\documentclass[12pt] {article}

\usepackage{amssymb}
\usepackage{epsfig12}

\def\ostrutdim{3ex} \def\ustrutdim{1ex}
\def\ostrut{\vrule height \ostrutdim width 0pt \relax}
\def\ustrut{\vrule width 0pt depth \ustrutdim \relax}
\def\oustrut{\ostrut\ustrut}  

\def\gev{$\mbox{GeV}^2$}
\def\lqcd{$\Lambda_{\mathrm{QCD}}$}
\def\aps{\alpha_s}
\def\apsb{\overline{\alpha}_s}

\begin{document}

\setlength{\baselineskip}{0.75cm}
\setlength{\parskip}{0.45cm}
\begin{titlepage}
\begin{flushright}
\large
DO-TH 96/25\\ February 1997
\end{flushright}
\normalsize
\vspace{1.5cm}
\begin{center}
\LARGE
{\bf Limitations of small~$x$ resummation methods from $F_2$ data} \\
\vspace{1.5cm}
\large
I.\ Bojak and M.\ Ernst\\
\vspace{1cm}
Institut f\"ur Physik, Universit\"at Dortmund \\
D--44221 Dortmund, Germany
\end{center}
\vspace{2cm}
\underline{{\large{Abstract}}} \\[2ex]
\normalsize
\setlength{\baselineskip}{0.75cm}
\setlength{\parskip}{0.45cm}
We discuss several methods of calculating the DIS structure
functions $F_2(x,Q^2)$ based on BFKL-type small~$x$ resummations.
Taking into account new HERA data ranging down to small $x$ {\em and\/}
low $Q^2$, the pure leading order BFKL-based approach is excluded.
Other methods based on high energy factorization are closer to conventional
renormalization group equations. Despite several difficulties and
ambiguities in combining the renormalization group equations with
small~$x$ resummed terms, we find that a
fit to the current data is hardly feasible, since the data in the
low $Q^2$ region are not as steep as the BFKL formalism predicts. Thus we
conclude that
deviations from the (successful) renormalization group approach towards
summing up
logarithms in $1/x$ are disfavoured by experiment.
\end{titlepage}
\newpage
%
%
%
\section{Introduction}

Recent measurements of the proton's structure functions
\cite{HERAold,HERA93,HERA94} have raised the question whether the observed
rise of $F_2$ at small values of $x$
is due to resummation effects described by the BFKL equation \cite{bfkl}.
In contrast, this observed rise has already been predicted using
the conventional renormalization group equation (RGE)
\`a la Altarelli-Parisi \cite{ap} and evolving from a low starting scale
\cite{grv90,grv}. That even data at low $Q^2$ can be described by the RGE
is now widely accepted \cite{MRSR,CTEQ4}. So the problem remains if one can find
a unique signature in the current $F_2$--data for BFKL-- or RGE-based evolution
equations, or if there is even the possibility of combining the different
approaches, aiming at a {\em unified\/} evolution equation, which might lead
to a better description of the structure function data.

The first question has already been answered in \cite{oldpap}. An update with recent
data appeared in \cite{resum}. These calculations were based on the
approach of Askew et al.\ \cite{AKMS}, treating the BFKL-equation as an
evolution equation in $\ln 1/x$, and calculating the structure functions
from the resulting unintegrated gluon distribution via the $k_T$-factorization
theorem \cite{factheorem,catani}. With a consistent choice of the infrared parameters,
we obtained steep structure functions incompatible with the data.

In section \ref{ktfac} we have a look at the high energy factorization
that was developed by Catani et al.\ \cite{factheorem,catani,cfm1}. This corresponds to the
small~$x$ limit, since $s \gg Q^2$ means $x \sim Q^2/s \ll 1$.
Based on this work, Forshaw et al.\ \cite{forshaw} calculated $F_2$ and $F_L$
by only considering the gluon sector, and by only taking into account
small~$x$ data ($x<10^{-2}$). We will show the limitations of this method in
section \ref{FRT}.

In section \ref{EHW} we try to modify the conventional
RGE with the complete small~$x$ resummed anomalous dimensions of Catani
et al., which include also the quark sector in next-to-leading order.
In a first attempt Ellis et al.\ \cite{ehw} evolved existing parton distributions with
this approach. Bl\"umlein et al.\  \cite{blum} later focused on the implementation 
of energy-momentum conservation. Employing these ideas,
we will try to {\em fit} the current data with special emphasis on the low $Q^2$ region
while constraining the partons in the large $x$ region.
We supply the details of our previously published results \cite{resum} and extend the
analysis of the conserving factors.

Our results confirm earlier related work by Ball et al.\ \cite{bafo1,bafo2},
in which energy-momentum conservation was implemented differently. There
three small $x$ parameters of the input distributions were fitted to
experimental data and the large $x$ parameters were matched to
existing conventional parton distributions. 
We note that a recent resummed calculation by Thorne \cite{thorne} using
leading order ``physical anomalous dimensions'' \cite{catphys}
apparently comes to contrary results.
We will summarize our results in section \ref{general}, and comment on the
limitations of BFKL inspired calculations of $F_2$ at
relatively low momentum transfers $Q^2$.

\section{High energy $k_T$-factorization} \label{ktfac}

Catani et al.\ \cite{catani,cfm1} proposed a method of summing small~$x$
logarithms which is based on the high $k_T$-factorization. It
has the advantage of being compatible with the conventional collinear
factorization. We just want to present the basic outline, without going
into details. 
First, we define the renormalization group equations in Mellin space:
\begin{equation}
	\frac{\partial f_i(n,Q^2)}{\partial ´\ln Q^2}=\sum_j \gamma_{ij} f_j(n,Q^2).
\end{equation}
Our interest now focuses on the quark singlet
combination $f^s\equiv x\sum_{i=1}^{N_f} (q_i+\overline{q}_i)$ and the
gluon density $f^g\equiv xg$. Note that we have rescaled the parton
distribution by $x$ for convenience. This implies that the anomalous
dimensions are
connected to the $x$-space splitting functions $\mathcal{P}_{ij}$ via
\begin{equation}
	\gamma_{ij}(n)=\int_0^1 x^n\,\mathcal{P}_{ij}(x)\,dx
	=\mathcal{P}_{ij}(n+1),\label{melhin}
\end{equation}
which differs by one from the standard definition. These anomalous dimensions
are in conventional perturbation theory given as a power series
in $\aps$, of which the first two orders are known today.

With the method of Catani et al., we are now looking for corrections
to the anomalous dimensions and coefficient functions
of the form $\aps^{i+k}/n^k$. Up to now, these have been calculated
for $i=0$ to all orders in $\aps/n$, and partially for $i=1$.
The singularities in moment
space ($n\rightarrow 0$) correspond to the small~$x$ limit in $x$-space.

The corrections to the gluon anomalous dimension can be inferred from the
solution of the characteristic equation ($\apsb=C_A\aps/\pi$; for QCD
$C_A=3$, $C_F=4/3$, $T_R=1/2$):
\begin{equation}
	\apsb\chi(\gamma_L)=n;\quad\chi(\gamma)=2\psi(1)-\psi(\gamma)-
		\psi(1-\gamma) \label{chareq}.
\end{equation}
The gluonic small $x$ resummed anomalous dimensions can then be expressed as
\begin{equation}
	\gamma_{gg}=\gamma_L;\quad\gamma_{gq}=\frac{C_F}{C_A}\gamma_L.
\end{equation}

The quark sector only contributes in the next-to-leading order, introducing
an additional factorization scheme dependence in the expressions. A simple
way to find this contribution is to look at the physical observable $F_2$.
One can calculate
its scaling violations in $n$-moment space in the limit $n\rightarrow 0$
from the gluon distribution via
the following expression ($\langle e^2\rangle$ denotes the mean squared
electric charge of the involved quarks):
\begin{eqnarray}
	\label{defcg}
	\frac{\partial F_2}{\partial\ln Q^2}&=&\langle e^2\rangle
	\left(C_2^g\gamma_{gg}+2N_f\gamma_{qg}\right)f^g \\
	&=&\langle e^2\rangle h_2(\gamma_L)\,R_n(\gamma_L)\,f^g \label{deff2h}
\end{eqnarray}
The function $h_2$ is basically the Mellin-transform of the off-shell
cross section $d\hat{\sigma}/d\ln Q^2$, whereas $R_n$ is a process-independent,
but (in general) factorization scheme dependent renormalization factor.
They are given by \cite{catani}
\begin{equation}
	h_2 (\gamma)
	= \frac{\aps}{2\pi}N_fT_R\frac{2(2+3\gamma-3\gamma^2)}{3-2\gamma}
	\left(\frac{\pi^2\gamma^2}{1-4\gamma^2}\frac{1}{\sin(\pi\gamma)
	\tan(\pi\gamma)}\right)
\end{equation}
and
\begin{eqnarray}
	R_n^{\overline{\mathrm{MS}}}(\gamma)&=&\left\{\frac{\Gamma(1-\gamma)\chi(\gamma)}{\Gamma
	(1+\gamma)[-\gamma\chi'(\gamma)]}\right\}^{\frac{1}{2}} \times
	\nonumber \\
	& & \times\exp\left\{\gamma\chi(1)+\int_0^\gamma d\gamma
	\frac{\psi'(1)-\psi'(1-\gamma)}{\chi(\gamma)}\right\}. \label{defrn}
\end{eqnarray}

We now use the DIS factorization scheme, which is defined by the requirement
for the Wilson coefficients $C_2^g=0$, $C_2^q=1$. Then the quark anomalous
dimension simply becomes
\begin{equation}
\label{DISres}
	2N_f\gamma_{qg}^{\mathrm{DIS}}=h_2(\gamma_L)R_n^{\mathrm{DIS}}(\gamma_L)
\end{equation}
(cf. Eq.~(\ref{defcg})). Fortunately, the expression for $R_n^{\mathrm{DIS}}$ remains the
same (Eq.~(\ref{defrn})), and as $\gamma_L$ is per definition scheme
independent, we can obtain fully resummed expressions in the gluon {\em and\/}
quark sector.
We get $\gamma_{qq}$ from the following relation ($\gamma_{NL}\equiv
\gamma_{qg}$):
\begin{equation}
	\gamma_{qq}=\frac{C_F}{C_A}\left(\gamma_{NL}-
	\frac{\aps}{2\pi}T_R\frac{2}{3}\right).
\end{equation}
Explicit $\overline{\alpha}_s /n$ series expansions of the used terms can
be found for example in \cite{cfm1,bafo1}.

With these corrections $\gamma_L$, $\gamma_{NL}$ the singlet
small~$x$ anomalous dimension matrix can then be written as follows:
\begin{equation}
        \hat{\gamma}=\left(\begin{array}{cc} 0 & 0 \\
                         \frac{C_F}{C_A} \gamma_L & \gamma_L
                     \end{array}\right)+
                     \left(\begin{array}{cc}
                         2N_f\frac{C_F}{C_A}(\gamma_{NL}-
			 \frac{\aps}{2\pi} T_R \frac{2}{3}) & 2N_f
			  \gamma_{NL} \\
                         \gamma_\delta & \gamma_\eta
                     \end{array}\right)+
                     \mathcal{O}\left(\aps^2\left(\frac{\aps}{n}\right)^k
                     \right),\label{anommod}
\end{equation}
where $\gamma_\delta$, $\gamma_\eta$ denote the yet unknown next-to-leading
order gluonic contributions. Without them we will have a residual scheme 
dependence in the physical observables. Calculations in the $Q_0$-scheme
\cite{q0a} suggest that the effects of setting them to zero will be
comparatively small due to the dominance of $\gamma_{qg}$ and running
coupling effects \cite{q0b}.
Note that the non-singlet part of the
renormalization group equation is not affected by small~$x$ contributions
as singular as $\gamma_L$, $\gamma_{NL}$.

\section{Gluon based approach} \label{FRT}
After this introduction to the formalism, we will focus on its implications
on the description of the measured structure function data. Our first
example is the approach developed by Forshaw et al.\ in \cite{forshaw}.
The basic idea is to use {\em only\/} the LO small $x$ contributions
$\gamma_L$ in the anomalous dimension matrix (\ref{anommod}) and to neglect
the quark singlet:
\begin{equation}
	f^g(n,Q^2)=f^g(n,Q_0^2)\,
		\exp\left(\int_{Q_0^2}^{Q^2}\frac{dk^2}{k^2}\gamma_L\right).
\end{equation}

Following the argumentation of the previous section, we can include
the renormalization factor $R_n$ into the gluon distribution before
calculating structure functions:
\begin{equation}
	G(n,Q^2)=R_n f^g(n,Q^2)=R_n f^g(n,Q_0^2) \exp[Z_n(Q^2,Q_0^2)]  \label{frtsol} 
\end{equation}
with
\begin{equation}
	Z_n(Q^2,Q_0^2)=\int_{Q_0^2}^{Q^2}\frac{dk^2}{k^2}\gamma_L.
\end{equation}

If we want to transform this back into $x$-space, we have to perform the
following complex integral:
\begin{equation}
	G(x,Q^2)=\frac{1}{2\pi i}\int_C dn\,x^{-n}R_n f^g(n,Q_0^2)
		\exp[Z_n(Q^2,Q_0^2)]. 
	\label{melinv}
\end{equation}
If now the gluon input $f^g(x,Q_0^2)$ is chosen as
a simple step function \cite{forshaw},
\begin{equation}
	f^g(x,Q_0^2)=\mathcal{N}\theta(x_0-x)
	\Longrightarrow f^g(n,Q_0^2)=\mathcal{N}\frac{x_0^n}{n},
\end{equation}
with a suitably chosen $x_0$, there is no singularity other than $n=0$, and
the integration contour in Eq.~(\ref{melinv}) can be chosen as a circle around 
the origin. Then it is possible to find an analytic solution to the integral
as a series of Bessel functions \cite{forshaw}.

In order to form the structure function $F_2$, one further integration
according to
Eqs.~(\ref{defcg}) and (\ref{deff2h}) has to be done.
Forshaw et al.\ proposed to include
the derivative of the coefficient function $\partial C_2^g/\partial\ln Q^2$
upon integration, although it
is formally subleading in Eq.~(\ref{defcg}), since
it may lead to contributions which are not negligible.
The missing input $F_2(x,Q_0^2)$ has been parametrized as $A+Bx^{-\lambda}$,
introducing three additional parameters. In \cite{forshaw}, $x_0$ was set
to $0.1$, and a (remarkably) low $\Lambda_{\mathrm{QCD}}^{(4)}$ of $115$~MeV
was chosen. This leaves five parameters
($Q_0^2$, $\mathcal{N}$, $A$, $B$, $\lambda$) free to be fitted against
structure function data. Using 1993 data \cite{HERA93} below $x=0.01$, they
obtain a very good $\chi^2$, which is slightly better when the
subleading terms mentioned above are included. It was our intention to test 
whether this 
approach still works for more recent data, which are more precise, and which
especially extend down to lower $Q^2$.

\subsection{Fits using the analytic formulae} \label{anfit}

Our first observation is that the excellent agreement with the data in
\cite{forshaw} does not last very long. The analytic formula implies that
in order to describe the low $Q^2$ data, one has to choose at least an equally
low $Q_0^2$. This already worsens the $\chi^2$ for the 1993 data
slightly (but it is still acceptable). If we now include the 1994 HERA data
and the E665 data \cite{E665}, the situation gets much worse, see table \ref{tabfrtfit}
for details. As in \cite{forshaw}, we set $x_0=0.1$, and only used data
with $x<0.01$.
\begin{table}
	\centering
	\begin{tabular}{|c|c||c|c|c|c|c|}
	\hline
\oustrut $Q_0^2$~[GeV$^2$] & data points & $\mathcal{N}$ & $A$ & $B$ 
	& $\lambda$ & $\chi^2$ \\
	\hline
	\hline
\oustrut $0.8$ & $257$ & $0.74$ & $0.21$ & $0.65\cdot 10^{-3}$ & $-0.94$ 
	& $650$ \\
	\cline{1-2}
\multicolumn{2}{|c||}{without subl.\ terms}\oustrut & $0.41$ & $0.42$
	& $-0.024$ & $-0.71$ & $900$ \\
	\hline
\oustrut $1.0$ & $250$ & $0.80$ & $0.23$ & $0.0018$ & $-0.79$ & $478$ \\
	\cline{1-2}
\multicolumn{2}{|c||}{without subl.\ terms}\oustrut & $0.51$ & $0.47$
	& $-0.0066$ & $-0.57$ & $646$ \\
	\hline
\oustrut $2.0$ & $232$ & $1.1$ & $-0.032$ & $0.12$ & $-0.26$ & $337$ \\
	\cline{1-2}
\multicolumn{2}{|c||}{without subl.\ terms}\oustrut & $0.71$ & $0.44$ 
	& $0.48\cdot 10^{-20}$ & $-4.68$ & $307$ \\
	\hline
	\hline
\oustrut $2.0$ & $232$ & $0.39$ & $-2.40$ & $ 2.02$ & $-0.076$ & $1502$ \\
	\hline
	\end{tabular}
	\caption{Parameters for various fits to $F_2$ data. In the last row
	$\Lambda_{\mathrm{QCD}}^{(4)}=177$~MeV.}
	\label{tabfrtfit}
\end{table}
One can see that the fits including the subleading terms are in principle
better than those without. Nevertheless, the $\chi^2$ of all these fits
is not very convincing. It is better, the higher we chose $Q_0^2$, so we
focus in the following on $Q_0^2=2$~GeV$^2$.

As is visible in Fig.~\ref{frtfig1}, the agreement with the data is not
very good. The inclusion of the subleading terms (solid line) leads to curves
which are ``bent'' down to the negative region for very small $x$.
It is obvious that for these values of $x$ this approach is not
reliable anymore, as was already seen in \cite{forshaw}, too.
With the 1993 data, this was not an issue, as there were no data in this
kinematic region. But with the new data, this becomes a problem.
Regarding the other curve without the subleading terms (dashed line),
agreement with data is better, but we see also that the influence
of $F_2(Q_0^2)$ is bigger: The gluon normalization $\mathcal{N}$ is
smaller, and the parameters $B$ and $\lambda$ acquire almost absurd
values.

The next question which arises concerns $\Lambda_{\mathrm{QCD}}$.
It is clear that this has a serious impact on the calculation, since
the value of $\alpha_s$ controls $\gamma_L$ and thus the steepness
of the gluon and the structure function. We tried to use a more realistic
value of $\Lambda_{\mathrm{QCD}}^{(4)}=177$~MeV \cite{CTEQ3}.
This leads to an even stronger disagreement
with the data. Moreover, relative to the other curves, the influence of the
input $F_2(x,Q_0^2)$ is stronger, $\mathcal{N}$ being even smaller.
This leads to the conclusion that the fitting procedure tries to reduce
the influence of the (BFKL driven) gluon. To summarize, this simple analytic
approach is disfavoured by the recent data. Note again that these data
can be well described by a conventional renormalization group analysis.

However, there remains one way out: If we abandon the simple step
function like input gluon density, then the situation may be improved.
It was noted in \cite{forshaw}, that the analytic formula could
be extended to more complex input densities. Unfortunately, we find
this not to be the case: In the first place, one has to choose
an ansatz whose Mellin transform has a {\em finite\/} number of $n$-plane
singularities. This means e.~g.\ for a standard gluon input of the form
 \begin{equation}
\label{proglu}
	f^g(x,Q_0^2)\sim x^\alpha(1-x)^\beta,\label{sans}
\end{equation}
which transforms into a Beta-function, that either $\alpha$ or $\beta$
have to be integer. While this is not too strong a constraint, the next
problem arises immediately: For the analytic inversion it is
essential that the input can be written as a power series in $1/n$.
However, the Laurent series stemming from Eq.~(\ref{sans}) does {\em not\/}
converge. Apart from that, one has to take into account that the expense
of performing a fit based on a series of Bessel functions increases
quickly, due to the nested sums involved and the greater number of
parameters. So the analytic calculation is only feasible for the most simple
form of the input gluon. We have chosen to simply resort to numerical
methods, see below, but a more sophisticated analytic treatment is also
possible \cite{bafo3}. 

\subsection{LO gluon fit} \label{FRTfit}

Before any definite conclusions can be drawn on the applicability of this
approach of Forshaw et al.\ \cite{forshaw}, we have to overcome the 
restrictions
mentioned above. We do this by transforming the moment space expression
according to Eq.~(\ref{melinv}) back to $x$-space numerically.
%
This allows
us to use a conventional form for the
gluon which in $x$-space can be written as
\begin{equation}
\label{glupar}
x\, g(x,Q_0^2) = N\, x^{-\lambda}\, (1+\eta\, x)\, (1-x)^\gamma.
\end{equation}
The resummed gluon is evolved to $Q^2$ by multiplying this
input distribution with $R_n$ and $Z_n(Q^2,Q_0^2)$ according to
Eq.~(\ref{frtsol}). 
An additional advantage of this method is that predictions for $F_2$ and
 $F_L$ are obtained simply by multiplying in $n$-space the appropriate
Wilson coefficients with the gluon solution before
transforming to $x$-space numerically.

We can now also exploit the fact that $Z_n$ can
be calculated analytically. First note that due to the LO BFKL
characteristic equation (\ref{chareq}) we can write the measure of the integral
as
\begin{equation}
d\ln Q^2 = \frac{4C_A}{\beta_0\, n}\frac{d\chi(\gamma_L)}{d\gamma_L}
\,\gamma_L,
\end{equation}
and thus we can rewrite the integral as \cite{collins}
\begin{equation}
Z_n(Q^2,Q_0^2)=\int_{\gamma_L(n, Q_0^2)}^{\gamma_L(n, Q^2)}\, d\gamma\,
\gamma\, \frac{d}{d\gamma}\, \chi (\gamma )
=\left.\gamma_L\left[2\,\gamma_E+\frac{n}{\overline{\alpha}_s}\right]+\ln
\frac{\Gamma (\gamma_L)}{\Gamma (1-\gamma_L)}\right|_{Q_0^2}^{Q^2}.
\end{equation}
This equation can be evaluated numerically once $\gamma_L$
is known.

We have tested a large set of fits using the gluon parametrization
(\ref{glupar}) shown above. The starting scale $Q_0^2$ was also fitted,
but usually this parameter ended up at the upper limit set to the
lowest $Q^2$ of the fitted points. We varied this lowest $Q^2$ from $2.0$
\gev\ to $3.5$~\gev . Also we tested the effects of including the
subleading term, of constraining the small~$x$ growth of the background
 $F_2\sim x^{-0.08}$ and of constraining the gluon power
 $xg\sim x^{-\lambda}$ to $0<\lambda <2$. The results of the
fits to the same data as in the previous section are summarized
in Table~\ref{tabfrt}.
\begin{table}
\begin{tabular}{|c|c|c||c|c|c|c|c|}\hline
\oustrut $\frac{dC_g}{d\ln Q^2}$&$F_2^{bckgr.}$&$xg\sim x^{-\lambda}$
&\multicolumn{5}{|c|}{$\chi^2/d.o.f.$ for $Q^2\ge$}\\
\oustrut added&$\sim x^{-0.08}$&$0<\lambda <2$
&2.0~\gev &2.5~\gev &2.8~\gev &3.0~\gev &3.5~\gev\\\hline
\hline
\oustrut $ $&$ $&$ $&3.08&1.72&1.35&1.23&1.00\\\hline
\oustrut $ $&$ $&$\bullet$&3.12&1.77&1.41&1.26&1.04\\\hline
\oustrut $ $&$\bullet$&$ $&3.17&1.74&1.43&1.26&1.03\\\hline
\oustrut $ $&$\bullet$&$\bullet$&3.19&1.78&1.48&1.29&1.06\\\hline
\oustrut $\bullet$&$ $&$ $&2.30&1.02&.679&.625&.583\\\hline
\oustrut $\bullet$&$ $&$\bullet$&2.30&1.02&.680&.626&.583\\\hline
\oustrut $\bullet$&$\bullet$&$ $&2.43&1.33&1.00&.814&.689\\\hline
\oustrut $\bullet$&$\bullet$&$\bullet$&2.98&1.33&1.00&.814&.689\\\hline
\end{tabular}
\caption{Fits using the formalism of Forshaw et al.\ and the general
form of the gluon (\ref{glupar}). The left part of the table shows
the type of fit performed and the right part shows the $\chi^2$ per
degree of freedom for different lowest $Q^2$ of the fitted data.}
\label{tabfrt}
\end{table}

The fits including the subleading terms $dC_g/d\ln Q^2$ describe
the data much better. In spite of this we do not believe that these
terms should be included, since they always lead to an obviously
wrong behaviour in the small~$x$ region at low to medium $Q^2$.
A typical example is given by the dot-dashed curve in Fig.~\ref{frtgen}
that also displays other fits of Table~\ref{tabfrt}. It shows
the $2.8$~\gev\ fit of the bottom row in Table~\ref{tabfrt}. The negative
contribution of these terms overwhelms the growth of $F_2$ in
the region $\sim 3-10$~\gev\ at small~$x$. The data constrain the
fit, but it starts to fall right after the data point smallest in $x$. 
So the negative contribution on the one hand improves the description
of the existing data due to a slowed growth, but on the other hand
renders any prediction for future data at smaller $x$ impossible.

All the other curves in Fig.~\ref{frtgen} correspond to fits without
subleading terms, but constrained in $F_2$ and $xg$ (corresponding to row four
in Table~\ref{tabfrt}). For all fits except for the one with
 $Q^2\ge3.5$~\gev, the fitted $Q_0^2$ equals the lowest $Q^2$ bin,
so that the corresponding curve in that bin shows the $F_2$
background. The fitted background of the exception is shown
separately in the $3.5$~\gev\ bin.
   
A general trend is that the fits get worse as lower $Q^2$ data
are included. This is primarily due to the strong growth induced
by the BFKL resummed anomalous dimension at low $Q^2$. 
As is obvious by comparing the solid curves
at $2.0$~\gev\ and at $2.8$~\gev\ in Fig.~\ref{frtgen}, the growth
at small~$x$ is enormous over a short evolution length. Thus the
fit forces the background to fall towards smaller $x$ to be able
to describe data at higher $Q^2$ at all. To a lesser extent this
is also true for the fits starting at $2.8$ and $3.5$~\gev .
The fit stays below the data at low $Q^2$ and small~$x$ in order
to describe the bulk of data at higher $Q^2$. Above $12$~\gev\
all fits describe the current small~$x$ data well. But if
the precision and depth in $x$ of the HERA data continues to
improve, this limit may go up.

We can conclude that despite some improvement upon using a more general
gluon input, the fits still cannot describe the data well if started from low $Q^2$
scales. Even though the subleading terms lower the $\chi^2$, a closer look
reveals that their influence at small~$x$ leads to unphysical results.
A word of caution has to be said about the influence of \lqcd . Lowering
it to values as low as the one considered in the original paper \cite{forshaw}
can halve the $\chi^2$ of the fit, since the effects of the low starting scales
are compensated somewhat. But in adopting such a low value of \lqcd\
one basically loses the connection to the large $x$ region completely.
If one attempts to vary \lqcd\ in fits from low starting scales $Q_0^2$, then one sees
soon that \lqcd\ always drops to the lowest limit set. This would of course
not happen if there were large $x$ data constraining the fit. So we here have
chosen to fix it at a realistic value instead.

\section{Inclusion of the quark sector} \label{EHW}

We have seen that all attempts to describe the structure function data
using only a small $x$ modified gluon density have failed.
Thus the next logical step is to incorporate the more or less neglected
quark densities.
At the same time, we want to stay as close to the successful
conventional renormalization group
equations as possible.
A method to achieve this was suggested by Catani et al.\ \cite{catani}
and used in the paper of Ellis et al.\ \cite{ehw},
but only by evolving an existing set of parton distributions
(MRSD$_0'$, \cite{mrsd0p}). In the following, we describe this
approach briefly and present the results of our fits based on this method.

The basic idea is to take the renormalization group equations in their
two-loop-form, and to modify the anomalous dimension matrix according to
Eq.~(\ref{anommod}). 
The importance of implementing the fundamental energy-momentum
conservation for the modified equations has already been stressed in \cite{ehw}.
Energy-momentum is of course always conserved in the conventional
formalism. 
In Mellin space this means that the first moments of the anomalous dimensions
have to vanish:
\begin{equation}
	\sum_i\gamma_{ij}(n=1)=0. \label{momsum}
\end{equation}
The all order small~$x$ resummations violate this equation and it has to be
enforced by hand somehow.
Additionally, when we combine the two-loop with the small~$x$
expressions, we have to avoid double counting of the leading terms which
appear in both the two-loop expressions and the resummed expressions.
This can easily be done by simply subtracting the first term in $\gamma_L$,
and the first two terms in $\gamma_{NL}$. 
As we are now working in next-to-leading order, we have to choose a
definite factorization scheme.
In the following we will work in the DIS factorization scheme, since
in this scheme there exists a fully resummed expression for $\gamma_{NL}$, as
mentioned earlier. 

Concerning energy-momentum conservation, several ansaetze have been
suggested
in the literature. Ellis et al.\ \cite{ehw} used two different ones, a ``hard''
one where the small $x$ corrections $\gamma_{ij}(n)$ are replaced 
by $\gamma_{ij}(n)-\gamma_{ij}(1)$,
and a ``soft'' one where these corrections are multiplied with a
factor $(1-n)$. Bl\"umlein et al.\ \cite{blum} additionally proposed the
``conserving factors'' $(1-n)^2$ and $(1-2n+n^3)$. Although these are
all arbritrary implementations of the momentum sum rule, one can study with
these different factors
the possible impact of yet unknown higher order 
corrections\footnote{Ball and Forte \cite{bafo1} have circumvented the 
energy-momentum conservation problem
in a different way. They used a factorization scheme in which
the unknown higher order terms $\gamma_\delta$, $\gamma_\eta$ in the
anomalous dimension matrix are {\em defined\/} by the momentum sum rule
Eq.~(\ref{momsum}).}.
In \cite{blum} it was already shown that with these factors it is possible
to suppress the small $x$ corrections, or even to overcompensate
the expected growth at small $x$.

There are some theoretical problems associated with these ansaetze.
First, we consider the hard implementation. While this does not modify
the $n$-dependence of the resummed corrections at all, it requires
their evaluation at $n=1$. This causes a problem if the starting
scale of the evolution is low, since the resummed $n$-plane pole $n_r=\apsb\,4\ln 2$ can in this case
pass through $n=1$, making further calculations impossible.
For example, the MRS R1 fit \cite{MRSR} uses $\Lambda_\mathrm{QCD}=241$~MeV and
a starting scale $Q_0^2$ of 1~\gev , leading to $n_r = 1$
at  $Q^2=1.07$~GeV$^2$ using the two-loop formula for $\aps$.
Moreover, the generated
structure functions are still much too step to describe the data, so we
will not use it further.

Second, factors involving higher powers of $n$
than $n^2$ (like one of those in \cite{blum}) cause a factorization problem.
If we remember that the leading term of $\gamma_{qg}$ beyond the two-loop
expression is proportional to $(\apsb/n)^2$, such
conserving factors imply that $\gamma_{NL}$ times the conserving factor
does not vanish anymore
for $n\rightarrow\infty$. This means it cannot be inverted to $x$-space
{\em on its own}. 
Since the anomalous dimensions
are folded with parton distributions which for reasonable choices of the
input fall off better than $1/n$ for $n\rightarrow\infty$, this is not a
problem for the evolution program, but it is still an indication that 
this product is not correctly factorized anymore. For this reason we do
not examine even higher powers of $n$ in the conserving factor.

Now we have to comment on our implementation of the resummed expressions
into the two-loop evolution program. Since we are using a program
which is based on the two-loop analytical solution in moment space of the renormalization
group equation \cite{grv90,furm}, this is not as simple as it would be 
e.~g.\ in
a Runge-Kutta based $x$-space evolution program. Let us first write down the singlet
renormalization group equation in the two-loop form including the small $x$
resummations:
\begin{equation}
	\frac{d\vec{f}(n,Q^2)}{d\aps}=\left[\frac{\aps}{2\pi}
	\hat{\gamma}^{(0)}(n)+\left(\frac{\aps}{2\pi}\right)^2
	\hat{\gamma}^{(1)}(n)+\hat{\gamma}^{\mathrm{res}}(n,\aps )
	\right]
	\frac{1}{-\frac{\beta_0}{4\pi}\aps^2-\frac{\beta_1}{(4\pi)^2}\aps^3}
	\vec{f}(n,Q^2)
\label{twoleq}
\end{equation}
The vector $\vec{f}$ consists of the quark singlet $f^s$ and the gluon
density $f^g$, and we employed the two-loop approximate solution for $\aps$:
\begin{equation}
	\frac{\aps(Q^2)}{4\pi}=\frac{1}{\beta_0\ln (Q^2/\Lambda^2)}
	-\frac{\beta_1\ln\ln(Q^2/\Lambda^2)}{\beta_0^3[\ln(Q^2/\Lambda^2)]^2}.
\end{equation}

The solution of Eq.~(\ref{twoleq}) is not trivial, since we are dealing
with a matrix equation; furthermore, it is unclear how to incorporate
the resummed expressions involving higher powers of $\aps$ into an
approximate solution which is accurate only to $\mathcal{O}(\aps^2)$. We do this
by separating the treatment of higher order terms $\aps^i$, $i\ge 3$ from
the resummed terms $\aps\,(\aps/n)^k$, $k\ge 2$:
\begin{eqnarray}
\vec{f}(n,Q^2)&=&
\exp\Bigg[-\frac{2}{\beta_0}\,\hat{\gamma}^{(0)}(n)\,\ln\frac{\aps (Q^2)}{\aps (Q_0^2)}\nonumber \\
&&\quad -\frac{1}{\pi\,\beta_0}\,\left(\aps (Q^2)-\aps (Q_0^2)\right)\Big(\hat{\gamma}^{(1)}(n)-\frac{\beta_1}{2\,\beta_0}\,\hat{\gamma}^{(0)}(n)\Big) \\
&&\quad -\int_{\aps (Q_0^2)}^{\aps (Q^2)}\, d\alpha\, \frac{\hat{\gamma}
^{\mathrm{res}}(n,\aps )}
{\frac{\beta_0}{4\pi}\aps^2+\frac{\beta_1}{(4\pi)^2}\aps^3}+{\cal O}\left(\aps^2, 
\aps^2\,\left(\frac{\aps}{n}\right)^k\right)\Bigg]\nonumber ,
\end{eqnarray}
with $k\ge 2$. 

Since $\hat{\gamma}^{\mathrm{res}}$ starts with $\aps\,(\aps /n)^2$
after subtraction to avoid double counting, the leading power of
the integral is $(\aps /n)^2$. In expanding the matrix exponential
we would encounter commutators of all three terms. But the commutator
of the conventional NLO term with the resummation integral is suppressed
by $\aps$. This contribution would be of the same order as the
commutator of the {\em error} of the resummation with the LO RG
term and can be ignored. Thus we can attach all the resummed terms to
 $\hat{\gamma}^{(1)}$ via
\begin{equation}
	\hat{\gamma}^{(1,\mathrm{res})}(n,\aps )=\hat{\gamma}^{(1)}(n)
	+\frac{\pi\beta_0}{\aps(Q^2)-\aps(Q_0^2)}
	\int_{\aps(Q_0^2)}^{\aps(Q^2)}d\alpha\,
	\frac{\hat{\gamma}^{\mathrm{res}}(n,\alpha)}{\frac{\beta_0}{4\pi}
	\alpha^2+\frac{\beta_1}{(4\pi)^2}\alpha^3}.
\end{equation}
For the final solution we then treat the resummed anomalous
dimension $\hat{\gamma}^{(1,\mathrm{res})}$ exactly 
like the original $\hat{\gamma}^{(1)}$.
The exponential of the matrices is expanded and terms of
 ${\cal O}(\aps^2)$ not stemming from the resummation are dropped. 
Thus, concerning the resummed terms, the solution is accurate up to
 ${\cal O}(\aps^2\, (\aps /n)^k)$ consistent with the error inherent to
the resummation itself.

We have checked that  using this procedure we can reproduce the
results of \cite{ehw}. But we want to improve on the results shown there
by on the one hand refitting the input parton distributions and on the other
hand by lowering the starting scale $Q_0^2$ to values used in
recent conventional RGE fits. We expect that lowering the
starting scale leads to the same kind of problems as encountered
in the previous section, namely a steep growth in the region
of small $x$. Thus it is necessary to refit the parton distributions
to test if one can compensate this growth.

In principle one should do a fit to all relevant data to constrain the
(presumably) universal new input parton densities and $\aps$ as much
as possible. Then the analysis would be truly competitive to
conventional RGE analyses like \cite{MRSR,CTEQ4}. But the small $x$
resummations for most of the processes are unknown. Furthermore
the computing time would be prohibitive due to the more complicated
expressions. Even if we just concentrate on $F_2$, the calculations
are already rather time consuming.

Thus we take the following approach: we use the optimal MRS R1
conventional RGE fit \cite{MRSR} as our starting point. The ansaetze for the
input parton distributions can be found there. We assume that at large $x$
the new parton distributions would be similar to the R1 ones, since
the resummation effects should be small at large $x$ and since there
are many experiments constraining the parton distributions in this region.
In practice we calculate the original R1 partons and $F_2$ at $x=0.05$,
 $0.15$, $0.4$ and compare them to the new fit using an artificial
error of one percent. This assumed small error is necessary to force the fit to
match the large $x$ region, since as we will see the small $x$ region
is not well described and would overwhelm any weak constraint
at large $x$. The valence quark distributions are not directly affected
by the small $x$ resummations. They could be influenced indirectly
by trading momentum with the quark sea, but they are well constrained by
experiments and negligible in the region of small $x$. In order to save
computing time we do not fit the parameters of the valence quark input,
but keep them fixed at the R1 values. We also keep the value of \lqcd\ 
fixed, since it is mainly constrained by experiments in the large $x$
region as discussed in \cite{MRSR}.

We allow for an additional degree of freedom by fitting
 $\lambda_\Delta\stackrel{\mathrm{MRS R}}{\equiv}-0.3$
of $x\Delta\equiv x(\overline{d}-\overline{u})\sim x^{-\lambda_\Delta}$.
The correct $u-d$ flavour symmetry breaking is taken care of by the
large $x$ constraints of the sea.
The glue and the sea are now fitted at $x<5\cdot 10^{-2}$ to the data.
Of course in this region the HERA data \cite{HERAold,HERA93,HERA94} 
are dominant. Fitting to the whole bulk of HERA data would again be too
time consuming. So instead we use the R1 fit, which describes HERA data
very well, to generate $F_2$ data points in the $x$ range that is covered
by HERA. The errors are adjusted to match those of the experiment.
Finally it should be mentioned that
the appropriate adjustments of the parton distributions due to
the scheme change $\overline{\mathrm{MS}}\rightarrow\mathrm{DIS}^{\mathrm{res}}$
have to be made\footnote{The superscript ``res'' merely signifies the 
inclusion of the small
$x$ resummations, not a change of the scheme definition.}.
This is done automatically at the starting scale $Q_0^2$,
so that only the input distributions are affected. Since we are mainly
interested in how the parameters are changed when the small $x$
resummations are switched on and since we keep the valence quarks
fixed, only the conventional Wilson coefficients are used for the
transformation. This does not limit the fits, but the obtained parameters
should not be used for a direct $\overline{\mathrm{MS}}^{\mathrm{res}}$
calculation.

Let us mention that in the previous fits by Ball and Forte \cite{bafo1,bafo2} 
only the small $x$ powers $\sim x^{-\lambda}$ of the glue and sea and the
starting scale $Q_0^2$ were treated as free parameters. To be precise, they
evolved several sets of existing parton distributions, which are compatible
with large $x$ data, to $Q_0^2$ using conventional two loop RGE. The original
ansatz was then refitted to these resulting partons. Finally a fit to the
{\em experimental} data using these new input parton densities was performed,
in which {\em only} the mentioned three parameters were allowed to vary.
The large $x$ behaviour of the partons can be reproduced
by rearranging the various large $x$ parameters, so that for
example the normalization can change. This does of course affect the small $x$
region as well, so that we instead find it necessary to fit the complete
ansatz. Also they introduce a ``reference
value'' $x_0$ above which all resummed expressions are switched off.
Although their motivation to separate the large from the small~$x$ region
is clear, we see no need in the formalism to introduce such an arbitrary
scale. Nevertheless, their conclusions are quite similar to ours, since in a
later paper \cite{bafo2} $x_0$ is found to be almost zero, thus excluding any
need for small~$x$ resummed expressions in the evolution equations.
So we will basically confirm their results with our different method, which
determines {\em all} parton parameters for the small $x$ modified evolution by
directly fitting to experimental data.
 
To see the influence of the resummation on the parton densities we show in Fig.~\ref{frtfig2}
the ratio of the resummed gluon and quark singlet to the standard one. As expected the
ratio goes to one for large $x$, but the expected slower convergence of the sea curves
is also clearly visible. For larger $Q^2$ the influence of the resummation is much
reduced. Most of the difference is picked up at low $Q^2$ values which is
understandable considering the moving resummed pole. 
To test the impact of yet unknown higher order contributions
as mimicked by our conserving factors, we are especially interested in the 
first term $\sim n$; but changing its coefficient requires the inclusion of
higher powers of $n$.
Comparing the $(1-2n+n^2)$ and $(1-2n+n^3)$ curves, we see that their influence is strong.
In the latter case, the partons are even {\em smaller\/} than the standard
ones and a fit using
this factor leads to an unacceptable {\em fall} of $F_2$ in the small $x$ region. Thus it is evident that present small $x$ resummations have no predictive
power.

On the other hand we have already mentioned that exactly the term $\sim n^3$
and all even higher powers spoil the transformation back to $x$ space. 
If we require the anomalous dimensions to be transformable, the maximum power
allowed for the conserving factor is $n^2$. In combination with the strong large $x$
constraints we then always get partons and $F_2$ larger than the ones obtained by conventional NLO evolution. We checked this by fitting the conserving factor
as well, i.~e.\ we fitted $a$ of $(1-a\cdot n+b\cdot n^2)$ with $b=a-1$.
The fit gives $a=1.2944$ and satisfies the large $x$ constraints well, but
the $(1-n)^2$ curve does better in the small $x$ region.
Actually if we {\em only} fit $a$ and leave all other
parameters at their original MRS R1 value, we get $a=1.2651$.
So for these values of $a$ the effect of the small $x$ resummations at large $x$
seems to be minimized.

Such large effects from subleading terms are caused by the resummed
BFKL pole. For $Q_0^2=1$~\gev\ it lies around $n_r\simeq 1$. So considering
a contour close to the resummed pole we see that the main contribution of
the resummed pole comes for regions of $n$ where the change $n^2\rightarrow n^3$
in the conserving factor means a considerable change of the subleading terms. 
Of course this depends on the starting scale $Q_0^2$ as well, for example at
 $Q_0^2=4$~\gev\ we have $n_r\simeq 0.7$. So on top of the change of the
resummation itself the relative importance of the subleading terms will change.
This arbitrariness is just a sign of our ignorance concerning the subleading
terms, but at least we can exclude powers of $n$ higher than two from the conserving
factor to avoid picking up terms that will not occur in the real subleading expressions.

The results for the partons translate directly to $F_2(x,Q^2)$ shown in
Fig.~\ref{ehw_f2}. As expected the resummed calculations always overshoot
the data at small $x$. The influence of the terms down by one power in $n$
is strong as can be seen by the spread in the curves. The ``dips'' in the
curves are somewhat artificial, basically they are due to the too strong
rise of the calculated $F_2$ compared to the data. Then $\chi^2$ is
minimized by adjusting the partons so that the curves are below the
data at larger $x$ first. Even the curve using the $(1-n)^2$ factor is too steep
at small $x$; this is even more true for the other curves. The curve
with the fitted conserving factor $a=1.2944$ mainly leads to an optimal large $x$
behaviour. It is interesting to note that if we leave out all large $x$ constraints
and fit the conserving factor, we get a good description $(\chi^2 / \mathrm{d.o.f.}\simeq 1)$
of the small $x$ data for $a=2$. But then at larger $x$ we are totally off the $F_2$
and parton constraints. Due to the coincidence that the fitted $a$ for the best 
small $x$ description is close to the $(1-n)^2$ we tested for the complete fit,
we can claim that the $(1-n)^2$ curve is basically the best if one
wants to describe {\em all} $F_2$ data but emphasizes the small $x$ region.

A problem of all fits has to be mentioned. If we leave \lqcd\ as a free parameter
it always drops to low values. Despite the still large errors
on \lqcd\ and despite the fact that in principle all determinations of $\aps$
should incorporate small $x$ resummations to be consistent,
we believe that $\Lambda_{\mathrm{QCD}}^{n_f=4}\simeq 200$ MeV should be
taken as typical for a ``low'' $\aps$. Experimentally low values of $\aps$ stem
from data at large $x$ \cite{PDG} where resummation effects should be small, so
we expect that this holds true in the cases considered here. The effect of lowering
\lqcd\ is of course greatest for the largest resummation effects, so we show
a curve with the conserving factor $(1-n)$ and
 $\Lambda_{\mathrm{QCD}}^{n_f=4}=200$ MeV. The $\chi^2/$d.o.f. at
small $x$ is for the lower \lqcd\  reduced from 21 to 11.
A straight fit would yield $\Lambda_{\mathrm{QCD}}^{n_f=4}=180$ MeV, but
this value should not be taken too seriously, since it depends on how strictly
we implement the large $x$ constraints.

\begin{table}
\begin{center}
\begin{tabular}{|cc|c|c|c|}\hline
\multicolumn{2}{|c|}{$x\:[g|S|\Delta](x,Q_0^2)=$}&\oustrut RGE&\multicolumn{2}{c|}{Resummed}\\
\multicolumn{2}{|c|}{$Ax^{-\lambda}(1-x)^{\eta}(1+\epsilon\sqrt{x}+\gamma x)$} &\oustrut MRS R1&$(1-n)^2$&$a=1.2944$\\\hline
\hline
\oustrut &$(A_g)$&24.5&0.94777&36.09\\
\oustrut &$\lambda_g$&-0.41&0.26536&-0.547\\
\oustrut Gluon&$\eta_g$&6.54&3.9801&6.866\\
\oustrut &$\epsilon_g$&-4.64&-1.5297&-4.483\\
\oustrut &$\gamma_g$&6.55&4.2279&6.135\\
\hline
\oustrut &$A_S$&0.42&0.41647&4.801\\
\oustrut &$\lambda_S$&0.14&0.036837&-0.3928\\
\oustrut &$\eta_S$&9.04&10.873&8.815\\
\oustrut Sea&$\epsilon_S$&1.11&5.1104&-0.330\\
\oustrut $(\epsilon_\Delta =0,\eta_\Delta=\eta_S)$&$\gamma_S$&15.5&22.073&0.441\\
\oustrut &$A_\Delta$&0.39&0.084878&0.001299\\
\oustrut &$\lambda_\Delta$&-0.30&-0.59879&-0.1086\\
\oustrut &$\gamma_\Delta$&64.9&72.461&1473\\
\hline
\multicolumn{2}{|c|}{$\frac{\chi^2}{\mathrm{d.o.f.}}(F_2)$, $x<5\cdot 10^{-2}$}&\oustrut&1.96&12.0\\
\multicolumn{2}{|c|}{$\frac{\chi^2}{\mathrm{d.o.f.}}(F_2,\mathrm{partons})$,
 $x\ge 5\cdot 10^{-2}$}&\oustrut&126&6.04\\
\hline
\end{tabular}
\end{center}
\caption{The fitted partons using RGE plus small $x$ resummations in comparison with
the original RGE MRS R1 partons \cite{MRSR}. 
}
\label{tabpartons}
\end{table}
In Table~\ref{tabpartons} we show the parameters of those two fits that can be considered
optimal. $A_g$ is calculated via the momentum
sum rule and $\Lambda_{\mathrm{QCD}}^{n_f=4}=241$ MeV. $S\equiv 2(\bar{u}+\bar{d}+\bar{s})$ is the total sea quark distribution, and $\Delta\equiv\bar{d}-\bar{u}$; for further details and the valence distributions refer to \cite{MRSR}.
The $\chi^2$ for the $(1-n)^2$ conserving factor is dominated by the partons
constrained by an
artificial one percent error. A more realistic choice
of a five to ten times bigger error would of course reduce the $\chi^2$
considerably.
But we see that even at small $x$ the quality of the conventional RGE fit is
not matched. Most
of the $\chi^2$ comes here from the points with smallest $x$ at the lower $Q^2$ bins.
Looking at the parameters we see that the input sea has turned flat whereas the input gluon
is now growing with smaller $x$ instead of being valence like. The normalization
of the gluon is lowered correspondingly.

The input sea of the $a=1.2944$ fit has become valence like. The gluon has stayed more
or less the same. On the one hand we see that the fit cannot describe the small $x$ data at all.
On the other hand at large $x$ the MRS R1 partons and $F_2$ are reproduced well.
We can be sure that this fit would be compatible with experimental data at large $x$.
Due to the necessary artificial constraint at large $x$ it makes no sense to add
the $\chi^2$ for all $x$, so one cannot conclude that the $a=1.2944$ is the
overall better fit. A more sophisticated treatment of the large $x$ region would be
needed to determine a true best fit. But since there are still parts of the
NLO small $x$
resummation missing and since the conserving factors give just an estimate of
the influence of subleading terms, we feel that not much could be gained by
improving the large $x$ treatment for the time being.

In order to minimize the systematic error introduced by the missing terms,
we also employed the $Q_0$-scheme \cite{q0a}. 
As already mentioned, at least the $q\overline{q}$
contribution to the missing gluonic small $x$ anomalous dimensions is small
\cite{q0b} in this scheme. The Wilson
coefficients for $F_2$ and the small $x$ anomalous dimensions stay the same
as in the DIS scheme,
except for setting $R_n\rightarrow 1$ in Eq.~(\ref{DISres}). Since
$R_n=1+{\cal O}[(\overline{\alpha}_s/n)^3]$, no changes are introduced in
the conventional two loop expressions. The effect on $F_2$ of setting
$R_n\rightarrow 1$ is explored in Fig.~\ref{q0_f2f}. There the deviation
$(F_2^{Q_0}-F_2^{\mathrm{DIS}})/F_2^{\mathrm{DIS}}$ is displayed for different
$Q^2$ starting from the {\em same} input parton distributions.
The conserving factor
$(1-n)^2$ and the parameters of the corresponding fit of
Table~\ref{tabpartons} were used for the evolution in both schemes.

The deviation is less than ten percent in the $x$ and $Q^2$ range of the
data considered here. Also we can
see from Fig.~\ref{q0_f2f} that the naive expectation that the DIS sea
and thus $F_2$ should be larger at small $x$ due to $R_n>1$ only becomes
true for $x$ values at the edge of current data. We can even expect a 
$Q_0$-scheme fit to fare worse in the $x<5\cdot 10^{-2}$ range, since
the bulk of HERA data is in the region where the $Q_0$-scheme calculation
results in an even larger $F_2$. We have confirmed this expectation by
repeating the $(1-n)^2$ fit in the $Q_0$-scheme. Actually the growth of
$F_2$ in the main data range is enhanced by the strict large $x$ constraints.
This is shown by the dotted line in Fig.~\ref{q0_f2f}, which display the
deviation at 65~\gev\ for the partons {\em fitted} in the $Q_0$-scheme.
The strong large $x$ constraints force the deviation to zero at three points,
which leads to the notable, artificial oscillation in that region. Thus
the resulting large $\chi^2$ at small $x$ should only be taken as indication
that switching to the $Q_0$-scheme does not give us a better estimate of the
missing contributions at the moment. 

\section{General Discussion and Conclusions} \label{general}

Figure~\ref{slopes} presents a different look at our results. Here we show $\lambda\equiv
\partial\ln F_2 / \partial\ln (1/x)$ as determined by the H1 experiment \cite{HERA94}
in comparison with theory. To obtain $\lambda$ a fit of the type $C\, x^{-\lambda}$ is
made for $x<0.1$ for the H1 data. This should be understood as a cut, so that the
actual highest  and lowest $x$ value used depends on the data. The two conventional NLO
RGE calculations MRS R1 \cite{MRSR} and GRV '94 \cite{grv94} have been treated
by us in a similar way using the same highest $x$ values but extending the lowest
 $x$ to $10^{-5}$. We see the perfect match of the MRS R1 fit, and show for
comparison the dynamical GRV '94 \cite{grv94} predictions which, however, are sensitive
to the precise choice of the input scale.

We also show in Fig.~\ref{slopes} the slope predicted by the pure LO BFKL evolution. For this we
simply used the $\lambda$ of our parametrization \cite{oldpap}.
For the ``resummed'' calculations presented
in the last sections we again used the fitting method to obtain $\lambda$. But
to avoid the ``dips'' of the curves
presented in the last section we use a $x<10^{-3}$ cut and for the ones of
the section \ref{FRT} we even have to use $x<3\cdot 10^{-4}$. We show in
Fig.~\ref{slopes}  $\lambda$ for all starting scales $Q_0^2$ of the fits displayed in Fig.~\ref{frtgen} and for the different fits displayed in Fig.~\ref{ehw_f2}.

We see that the big spread in the predicted $\lambda$ for all
the BFKL inspired methods is diminishing towards higher $Q^2$.
{\em But} it lies always above the data except perhaps at very high $Q^2$.
So the resummed $F_2$ is always growing too fast in the small $x$ region,
since at very large $Q^2$ the conventional RGE terms are expected to become
dominant. This is also probably the explanation why the pure LO BFKL
$\lambda$ is the only one that keeps growing. There are no conventional
RGE terms that can take over at high $Q^2$ for this curve.
The fact that the LO resummed
calculations give predictions for $\lambda$ that are in the same ballpark as 
those including additionally the NLO resummations
can be attributed to the dominance of the BFKL pole.    

The failure of the pure LO BFKL formalism \cite{oldpap,resum} 
to describe the experimental $F_2$ data is not so
surprising. There has never been any guarantee that the resummation of only the leading logarithms
in $x$ would be appropriate in the kinematic region explored by current experiments.
The inclusion of a larger
part of the total contributions of perturbative QCD should na\"\i vely lead
to a better description of the data. Our calculations using
the methods of Forshaw et al.\ show us clearly that it is not sufficient
to use just gluons and the LO resummed anomalous dimensions in RGE
type calculations. The most promising method
should be to use all that is known: the NLO RGE and the LO and
part of the NLO resummations in $x$. In doing so we immediately encounter
the first hint of trouble. The resummations in $x$ always violate the fundamental
energy-momentum sum rule.

Thus we can really only rely on the calculations if the results are not
changed much when we implement this constraint in various ways.
In fact the results depend strongly on the implementation
and one cannot give reliable predictions for $F_2$ at 
small $x$ at all. The main effect stems from rather large $n$ and thus
variations in the conserving factor lead to changes that are formally
subleading but numerically important. If we limit our calculations to
conserving factors that do not introduce terms absent from higher order
anomalous dimensions, then we can still see a trend in the predictions:
again $F_2$ grows too strongly at small $x$.

Our results could be taken as an indication that the still missing NLO
resummation pieces will not be sufficient to improve the stability of the
calculation and its outcome.
This would prove that the small $x$ resummations do not lead to a
stable perturbative series in contrast to the RG calculations.
Of course this is entirely possible, but it would question
all future work on this subject. But there are some important caveats
concerning such a strong conclusion.

A strong growth that can not be suppressed by adjusting the parton distributions
is only encountered if the starting scale $Q_0^2$ is below approximately
3~\gev . To a lesser extent the variations induced by different conserving
factors increase for lower starting scales. Also the fits can always be
improved significantly by lowering \lqcd . Only a true fit to all the
large $x$ data would tell us if  \lqcd\ can be determined uniquely and if
the small $x$ region is well described with the value obtained. But our fits
already indicate that probably this would not
be the case. These problems are all connected to the strong BFKL pole, since
the fits always improve if $\aps$ becomes smaller and the pole moves to the
left in the $n$-plane.

It has to be mentioned as well that some closely related calculations do not
seem to encounter these problems. The ``physical anomalous
dimensions'' calculations \cite{thorne,catphys} relate only
physical observables $F_2$ and $F_L$ and thus any factorization scheme
dependence is avoided.
This fit to the data even seems to be preferred over conventional RG
calculations.
We note that due to the relative order $F_L\sim\aps F_2$
such calculations should be limited to leading logarithms until the
next-to-next-to-leading piece of the coefficient functions of $F_L$
is determined. This fit also
implies a rather low (LO, 4 flavours) \lqcd$=100$ MeV.  The input scale of
the evolution $Q_0^2$ is determined as 40~\gev\ and the fit becomes
uncompetitive for $Q_0^2\gtrsim 3$~\gev . Finally a new scale $A_{LL}$ is introduced
at which the ``inputs become nonperturbative'' \cite{thorne}.
The usual choice $A_{LL}=Q_0^2$ would simplify the formulae used considerably.
But this would spoil the agreement with data, since $Q_0^2$ would be low.

We suspect that if this kind of fit was redone starting from a low scale
$Q_0^2=A_{LL}$, then the same kind of problems as we have encountered would
occur.
Calculations using the colour dipole model are also giving
good agreement with recent $F_2$ HERA data \cite{navelet}, but note that they 
only used a {\em fixed\/} strong coupling. Another successful
approach \cite{kms} based on the CCFM equation \cite{cfm1,CCFM},
does not include the quark sector, uses approximations only valid at
small $x$ and represents a LO resummation. But the imposition of a purely
kinematic constraint on the gluon ladder introduced in \cite{kms} suggests
that higher order terms could suppress the growth in the gluon sector and
thus effectively shift the BFKL pole to the left.

For a final judgment on the small $x$ resummations we will have to
wait until the calculation of the NLO pieces will be completed. Also
it is probable that with a starting scale of around 4~\gev\ it will not
be possible to rule out such contributions. Only a true fit to
all the usual data using small $x$ resummations consistently will
tell us if the assumption that \lqcd\ can not be lowered too much
because of the large $x$ region is correct. Keeping this in mind,
we suggest that currently available methods and data strongly
favour conventional RG  calculations. If this statement survives the test
of time, it will mean that the small $x$ resummations do not
represent a good perturbative series. This would shed doubt on
all calculations involving them.

{\bf Acknowledgments:} We thank E.\ Reya for his advice and
A.\ Vogt for interesting discussions on the
conserving factors.
This work has been supported in part by the
'Bundesministerium f\"ur Bildung, Wissenschaft, Forschung und Technologie',
Bonn.

\newpage
%
%
%
\section*{Figure Captions}
\begin{figure}
\begin{center}
\epsfig{file=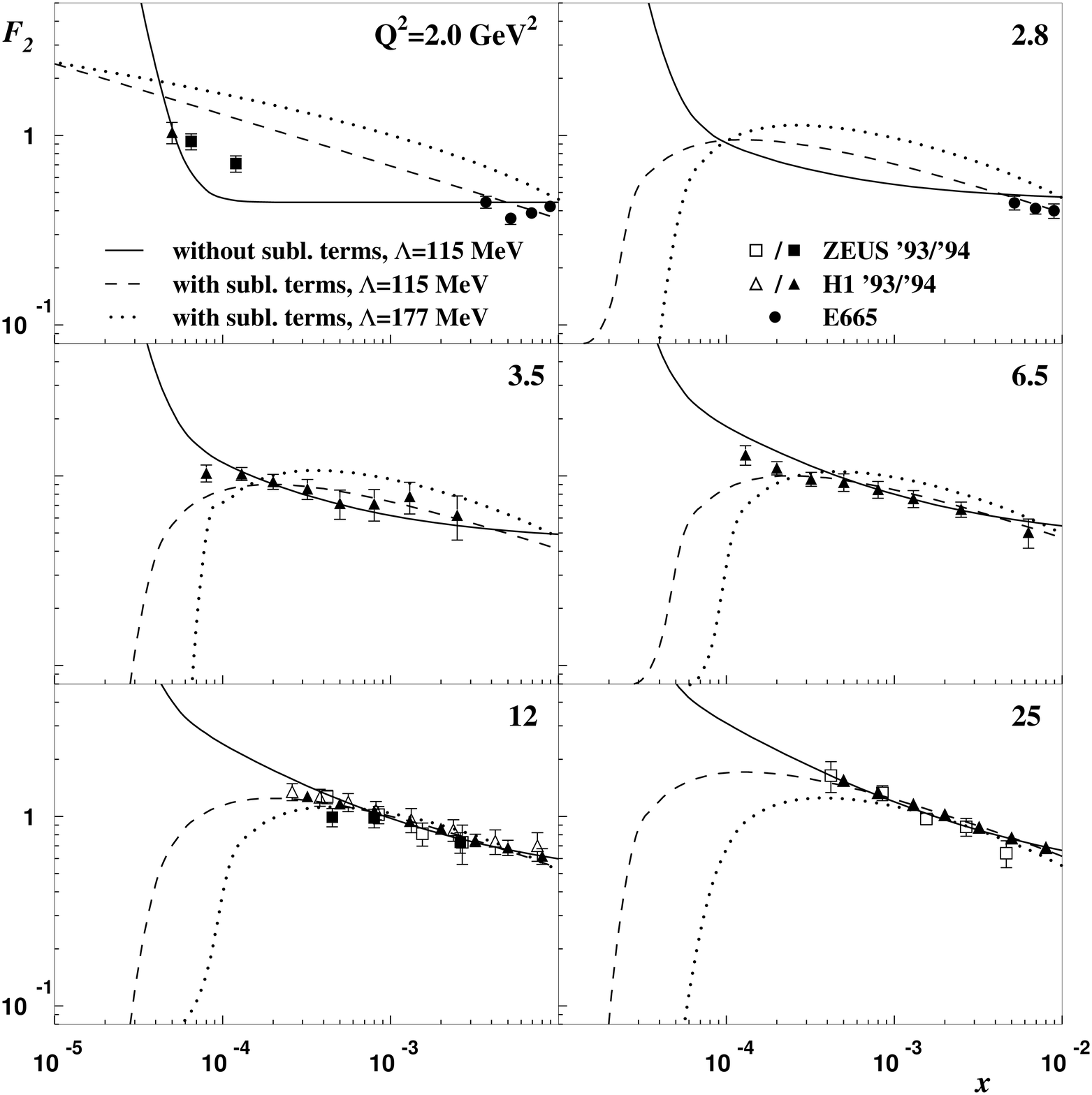,height=20cm,width=16cm}\\
\refstepcounter{figure}
\label{frtfig1}
\vspace{0.5cm}
{\large\bf Fig.\ \thefigure}
\end{center}
\end{figure}

\begin{figure}
\begin{center}
\epsfig{file=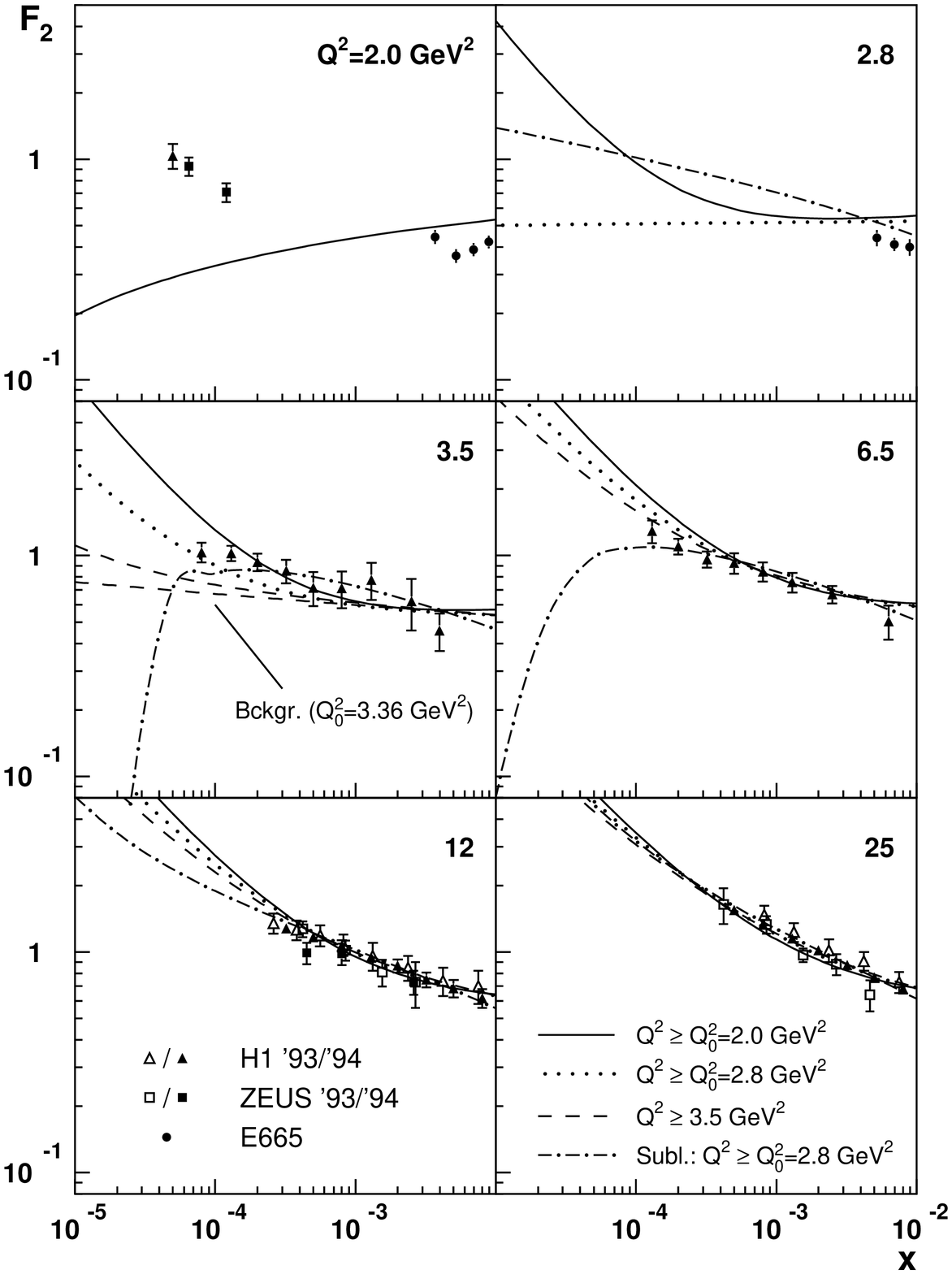,height=20cm,width=16cm}\\
\refstepcounter{figure}
\label{frtgen}
\vspace{0.5cm}
{\large\bf Fig.\ \thefigure}
\end{center}
\end{figure}

\begin{figure}
\begin{center}
\epsfig{file=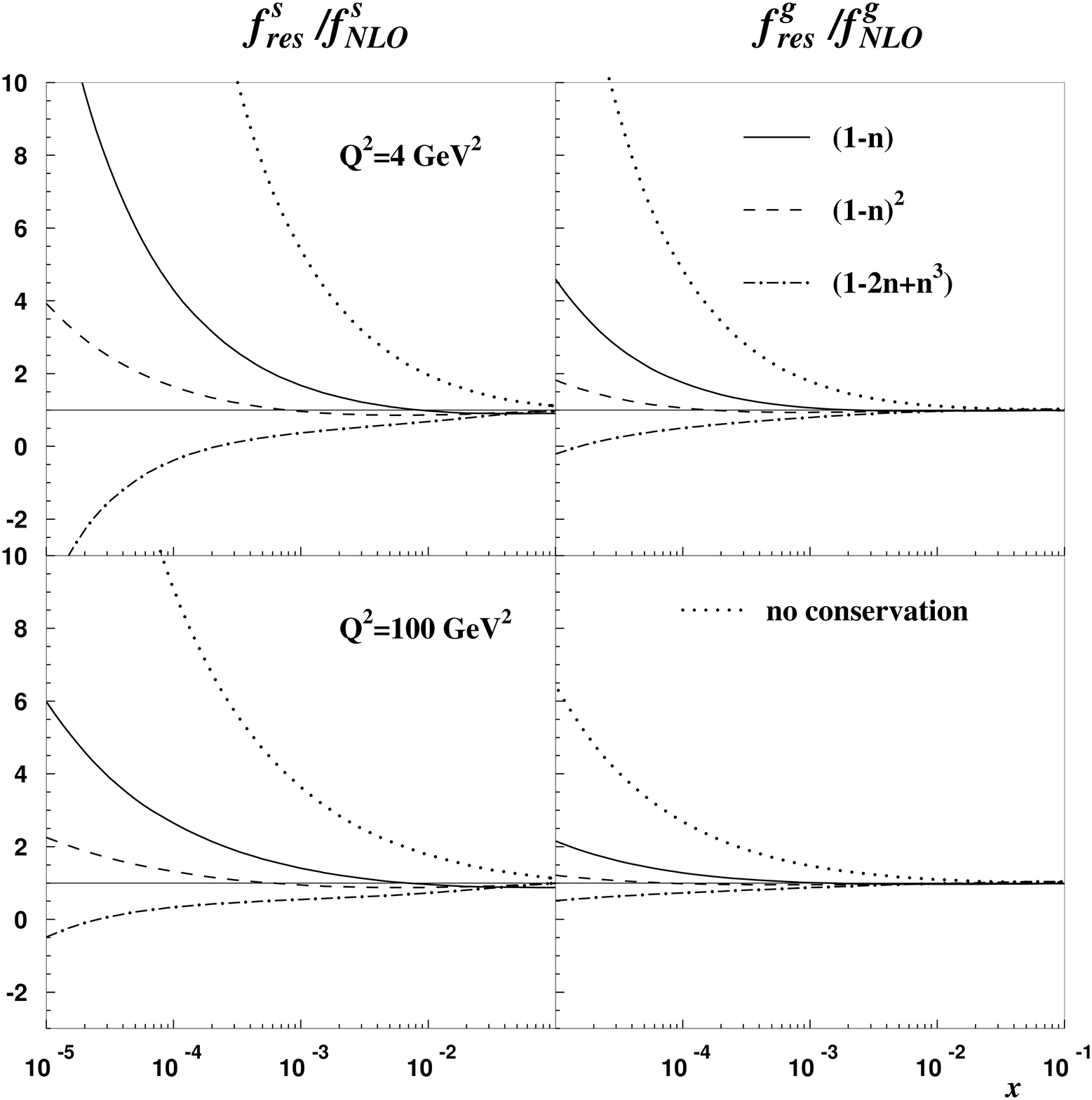,height=16cm,width=16cm}\\
\refstepcounter{figure}
\label{frtfig2}
\vspace{0.5cm}
{\large\bf Fig.\ \thefigure}
\end{center}
\end{figure}

\begin{figure}
\begin{center}
\epsfig{file=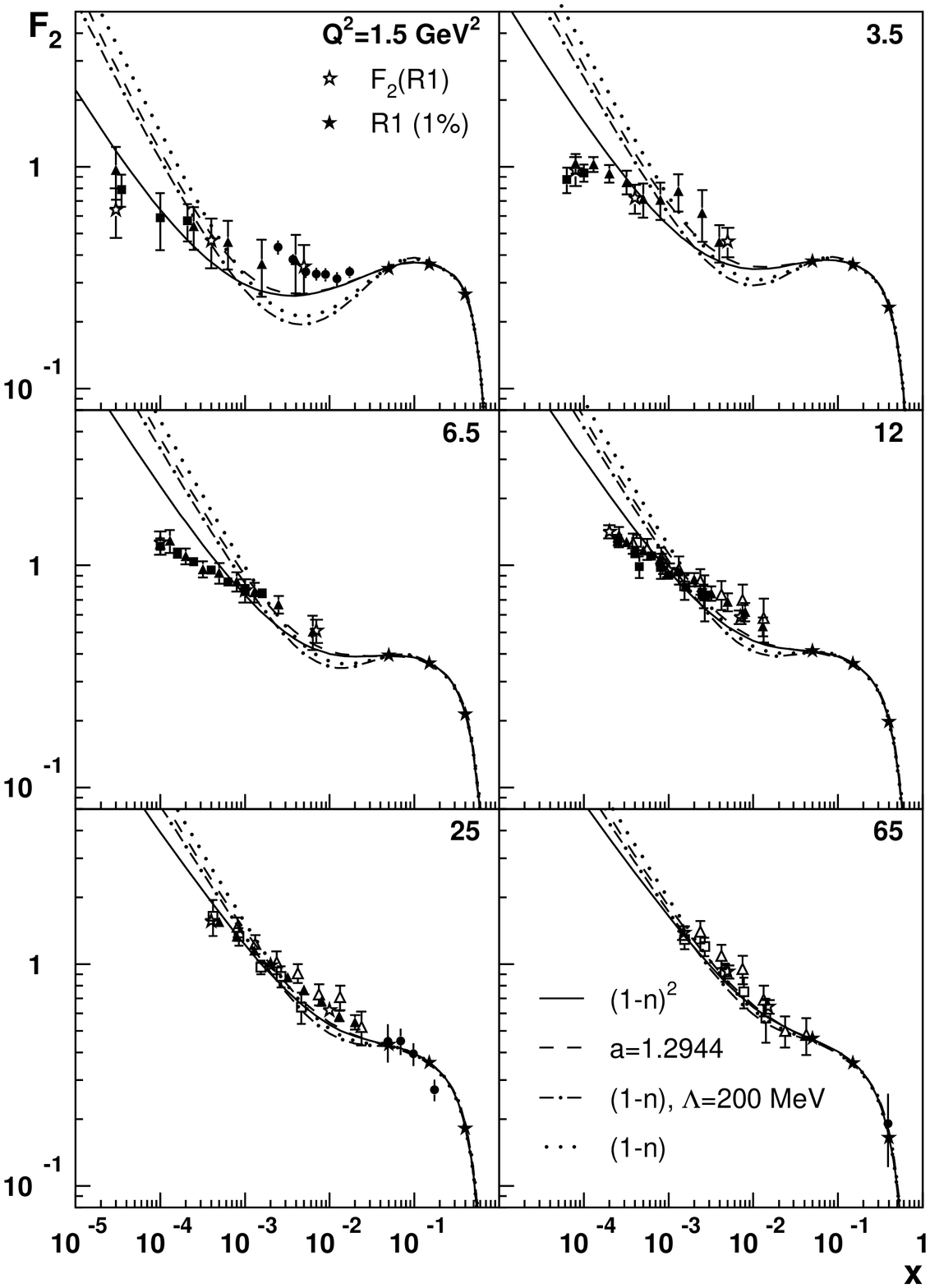,height=20cm,width=16cm}\\
\refstepcounter{figure}
\label{ehw_f2}
\vspace{0.5cm}
{\large\bf Fig.\ \thefigure}
\end{center}
\end{figure}

\begin{figure}
\begin{center}
\epsfig{file=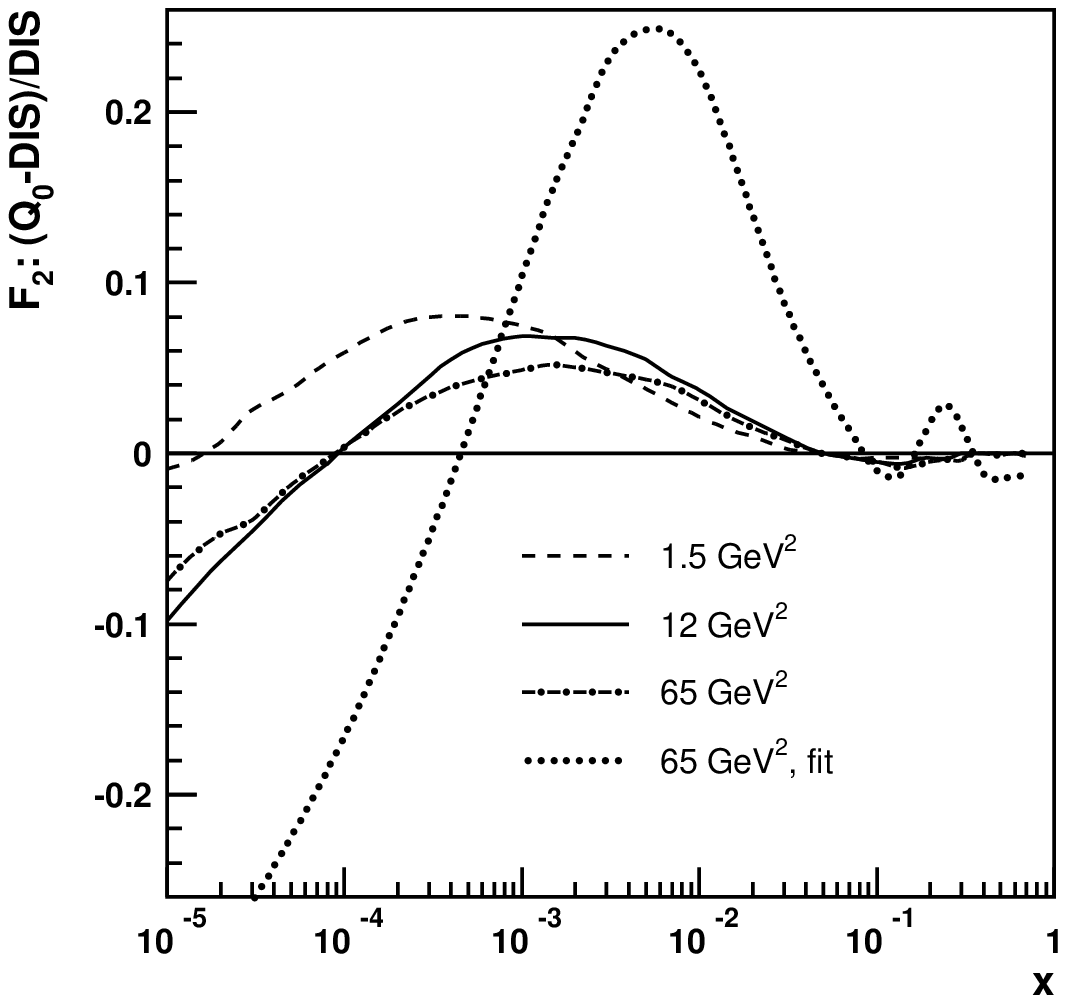,height=13cm,width=13cm}\\
\refstepcounter{figure}
\label{q0_f2f}
\vspace{0.5cm}
{\large\bf Fig.\ \thefigure}
\end{center}
\end{figure}

\begin{figure}
\begin{center}
\epsfig{file=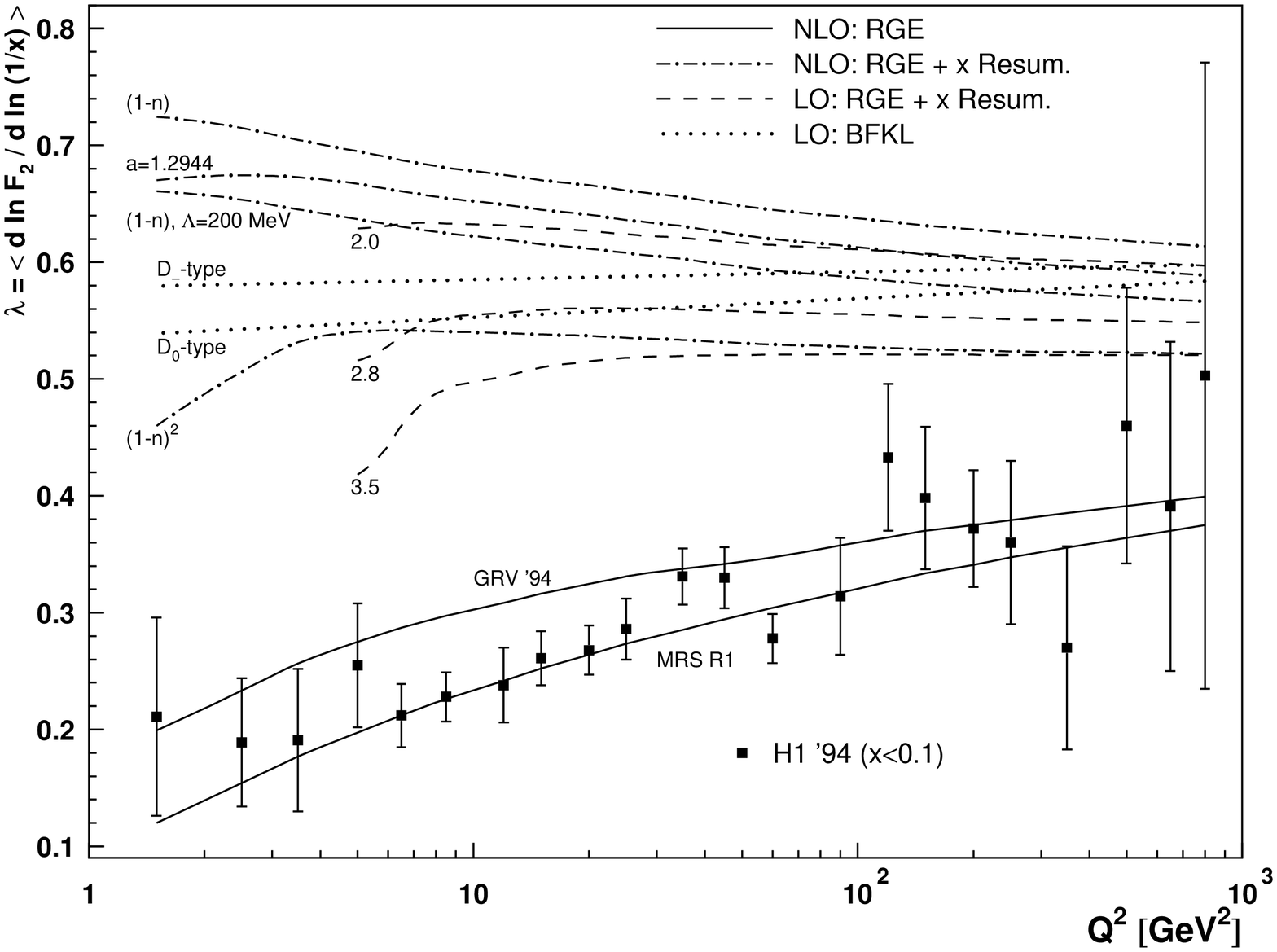,height=20cm,width=16cm,angle=90}\\
\refstepcounter{figure}
\label{slopes}
\vspace{0.5cm}
{\large\bf Fig.\ \thefigure}
\end{center}
\end{figure}

\begin{description}
\item[Fig.~\ref{frtfig1}] Structure function obtained with the analytic
prescription of Forshaw et al. \cite{forshaw}, with refitted parameters according to
the shown data from HERA \cite{HERA93,HERA94} and E665 \cite{E665}.
The solid curve is calculated without
the subleading term $\partial C_2^g/\partial\ln Q^2$, whereas the others curves
include them, with two different choices of \lqcd .

\item[Fig.~\ref{frtgen}] Fits to $F_2$ data as in Fig.~\ref{frtfig1} but using the
gluon in Eq.~(\ref{proglu}). The lower $Q^2$ cut of the fitted data is varied and
the fitted starting scale $Q_0^2$ ends up at this value except for the fit
 $Q^2\ge 3.5$\gev . For this fit the background at $Q_0^2=3.36$~\gev\
is displayed in the 3.5~\gev\ bin. The dot-dashed curve is an example
of a fit including subleading terms.

\item[Fig.~\ref{frtfig2}] Impact of the resummed terms in the evolution
equation on the singlet (left) and gluon part (right) relative to conventional
two-loop calculations. Also the influence of different methods to restore
energy-momentum conservation is shown. The unmodified MRS R1 distributions
\cite{MRSR} are used as input.

\item[Fig.~\ref{ehw_f2}] The results for $F_2$ of fits using
different conserving factors in a NLO calculation including
small $x$ resummations. The data used for the fit is shown as stars and experimental
data as in Fig.~\ref{frtfig1} is shown for comparison.

\item[Fig.~\ref{q0_f2f}] The deviation of $F_2$ calculated in the $Q_0$-scheme
\cite{q0a,q0b} from the DIS scheme at several $Q^2$. The parton
distributions of Table~\ref{tabpartons} obtained in the $(1-n)^2$ DIS scheme
fit are used as input. The dotted line shows the deviation at 65 \gev , when
partons fitted in the $Q_0$-scheme are used for $F_2^{Q_0}$.

\item[Fig.~\ref{slopes}] The slope of $F_2$ of the calculations
using small $x$ resummations compared with experimental data and
two conventional NLO RGE calculations. Results of the same method
are shown in one line style. The main parameter varied
to obtain the different curves of one method is displayed at the curve. 
\end{description}
%
%
%
\pagestyle{empty}
\vspace*{\fill}

\pagestyle{empty}
\vspace*{\fill}

\pagestyle{empty}
\vspace*{\fill}

\pagestyle{empty}
\vspace*{\fill}

\pagestyle{empty}
\vspace*{\fill}

\vspace*{\fill}
\end{document}